\documentclass{article}

\usepackage{arxiv}

\usepackage[utf8]{inputenc} 
\usepackage[T1]{fontenc}    
\usepackage{hyperref}       
\usepackage{url}            
\usepackage{booktabs}       
\usepackage{amsfonts}       
\usepackage{nicefrac}       
\usepackage{microtype}      
\usepackage{lipsum}
\usepackage{xspace}
\usepackage{cite}
\usepackage[numbers]{natbib}
\usepackage{xfrac}
\usepackage{caption}

\newcommand{\codename}{WristO\({_2}\)\xspace}
\newcommand{\spo}{\({SpO_{2}}\)}
\graphicspath{{images/}}

\title{\codename:~Reliable Peripheral Oxygen Saturation Readings from Wrist-Worn Pulse Oximeters}

\author{
  Caleb Phillips\\
  University of Toronto\\
  \texttt{caleb@cs.toronto.edu} \\
   \And
  Daniyal Liaqat\\
  University of Toronto and Vector Institute\\
  \texttt{dliaqat@cs.toronto.edu} \\
   \AND
  Moshe Gabel\\
  University of Toronto\\
  \texttt{mgabel@cs.toronto.edu} \\
   \And
  Eyal de Lara\\
  University of Toronto\\
  \texttt{delara@cs.toronto.edu} \\
}

\begin{document}
\maketitle

\begin{abstract}

Peripheral blood oxygen saturation ($SpO_2$) is a vital measure in healthcare.
Modern off-the-shelf wrist-worn devices, such as the Apple Watch, FitBit, and Samsung Gear, have an onboard sensor called a pulse oximeter. While pulse oximeters are capable of measuring both $SpO_2$ and heart rate, current wrist-worn devices use them only to determine heart rate, as $SpO_2$ measurements collected from the wrist are believed to be inaccurate.
Enabling oxygen saturation monitoring on wearable devices would make these devices tremendously more useful for health monitoring and open up new avenues of research.

To the best of our knowledge, we present the first study of the reliability of $SpO_2$ sensing from the wrist. Using a custom-built wrist-worn pulse oximeter, we find that existing algorithms designed for fingertip sensing are a poor match for this setting, and can lead to over $90\%$ of readings being inaccurate and unusable.
We further show that sensor placement and skin tone have a substantial effect on the measurement error, and must be considered when designing wrist-worn $SpO_2$ sensors and measurement algorithms.

Based on our findings, we propose \codename, an alternative approach for reliable $SpO_2$ sensing. By selectively pruning data, \codename achieves an order of magnitude reduction in error compared to existing algorithms, while still providing sufficiently frequent readings for continuous health monitoring.
\end{abstract}


\section{Introduction}\label{intro}

Peripheral oxygen saturation (\spo{}) has many uses in healthcare monitoring and is a primary vital sign used by nurses and physicians to monitor patients. It is a measure of the amount of oxygenated blood (expressed as a percentage) and it's usefulness extends across domains such as sleep apnea diagnosis~\cite{brouillette2000nocturnal}, monitoring oxygen therapy results for COPD patients~\cite{sliwinski1994adequacy}, and patient recovery monitoring in the ICU~\cite{taenzer2010impact}. However, most methods of \spo{} monitoring are intermittent and require active user interaction. For example, in hospitals, nurses often record \spo{} by attaching a fingertip device to patients during their rounds. At home, individuals concerned about their \spo{} level can purchase similar commercial devices and record their \spo{} a few times per day. As part of the growing mobile health monitoring movement, cell phone manufacturers have recently provided an onboard pulse oximeter on the back of smartphones (e.g Samsung Galaxy S8\footnotemark) that require the user to press a fingertip against the sensor to obtain an \spo{} reading. These devices all make use of a sensor called a pulse oximeter, which works by emitting light and measuring how much of the light is absorbed by the user's blood.

\footnotetext{https://www.samsung.com/global/galaxy/galaxy-s8/specs/}

Adding \spo{} measuring capabilities to wrist-worn devices seems like a logical next step. A significant advantage of monitoring \spo{} on the wrist is that because the device would be in constant contact with the user's skin, it eliminates the need for active interaction from the user and consequently, allows more frequent measurements. The ability to more frequency monitor oxygen saturation levels could provide a useful diagnostic tool, allowing for the development of early interventions that could drastically improve health outcomes and reduce health care costs. 

Interestingly, wrist-worn devices such as the Apple Watch, FitBit, and Samsung Gear already contain pulse oximeters. However they only use the data from the pulse oximeter to derive heart rate and not oxygen saturation. The pulse oximeters on these devices are fundamentally the same as the ones used in hospital and commercial fingertip \spo{} monitors, however calculating oxygen saturation from a wrist-worn sensor leads to mostly inaccurate and unreliable~\cite{mendelson2003measurement} data. This is primarily an issue of a poorly fitting devices, wrist/arm movement, low blood perfusion, and interference from ambient light.


Despite the fact that most pulse oximeter readings from a wrist-worn device are unreliable, it is our hypothesis that occasionally readings taken from such a device will be sufficiently reliable. Even if a small fraction of oxygen saturation readings are reliable, as long as they can be confidently identified among a majority of noisy readings, we believe that we can improve the current state of personal oxygen saturation monitoring. Consider a patient that currently tracks her oxygen saturation twice per day using an at home fingertip sensor kit. If she can use her smartwatch to identify a single reliable \spo{} reading every ten minutes, we have succeeded in increasing the amount of available data by almost two orders of magnitude. We have also removed the need for active user interaction.

In this work, we demonstrate that an intermittent reliable \spo{} signal can be taken automatically from the wrist using sensors similar to those currently employed in existing wrist-worn devices, such as the Apple Watch, FitBit, and Samsung Gear (given these devices employed a proper LED configuration). We develop a custom wrist-worn sensor collection platform and record data from ten participants. We implement a system, which we call \codename{}, that consists of a pipeline of automated feature extraction and a gradient boosting classifier to label signals as reliable or unreliable. \codename{} uses pulse oximeter and motion data to detect and reject unreliable data, which reduces the average error from 14.5\% to 1.5\% compared to a baseline implementation while generating a reading on average at least every three minutes. We also measure the effect of sensor placement and skin tone and show that \codename{} is robust to variations in skin tone. Furthermore, we show that \codename{} generalizes to unseen skin tones and participants and explore whether training participant specific models is beneficial.

The rest of this paper is organized as follows. Section~\ref{background} provides background on pulse oximeters, and shows why $SpO_2$ measurement from the wrist is challenging.
In Section~\ref{approach} describes our approach for building reliable wrist-worn pulse oximeters, and Section~\ref{implementation} details experimental setup and our data collection.
In Section~\ref{evaluation} we evaluate our approach and compare it to current algorithms.
In Section~\ref{future} we discuss practical deployment considerations.
Section~\ref{related} reviews related work, and Section~\ref{conclusions} summarizes.

\section{Background and Motivation}
\label{background}

Pulse oximeters are small sensors that allow non-invasive monitoring of heart-rate, blood oxygen saturation, and other health related metrics~\cite{mendelson1988noninvasive}.
A pulse oximeter consists of one or more light emitting diodes (LEDs, usually red and infrared), and a photodetector.
Light emitted by the LED interacts with the users blood and is then captured by the photodetector. 
Because oxygenated hemoglobin and non-oxygenated hemoglobin absorb different wavelengths of light, the amount of each wavelength of light captured by the photodetector indicates the level of oxygen in the blood.
This signal (called a \emph{photoplethysmogram}, or PPG), can be used to estimate heart rate, \spo{} and other metrics.

%
An estimate of oxygen saturation is produced from the PPG by calculating a ratio of ratios between the amount of red (660nm, absorbed mostly by non-oxygenated blood) and infrared (940nm, absorbed mostly by oxygenated blood) light detected, as described in Equation~\ref{spo2equation}:

\begin{equation}\label{spo2equation}
Sp{ O }_{ 2 }=y_{0}-m\times \left( \frac { \sfrac { { AC }_{ Red } }{ { DC }_{ Red } }  }{ \sfrac { { AC }_{ IR } }{ { DC }_{ IR } }  }  \right)
\end{equation}

\textit{AC} and \textit{DC} denote the alternating and direct current measured by the photodetector for each light source. These terms arise from the periodic nature of the cardiac cycle and the fact that the level of oxygenated arterial blood fluctuates with the cardiac cycle whereas the level of non-oxygenated blood in the veins stays fairly constant. The DC component reflects the periodic oxygenated blood while the AC component reflects the constant non-oxygenated blood. This, coupled with the fact that infrared is mostly absorbed by oxygenated blood means taking the ratio of ratios isolates the proportion of oxygenated blood in the artery.    

The currents for each reading are taken from a window of a fixed size, 4 seconds ($100$ samples at $25Hz$) in most cases. The \(y_{0}\) and \(m\) terms represent a linear fit for calibration, and would generally be provided by the manufacturer of the specific sensor after they have calibrated the sensor against a ground truth.
Equation~\ref{spo2equation} is based on Beer-Lamberts law, and described in more detail in \cite{mohan2016measurement}.

\subsection{Oxygen Saturation Extraction Algorithms}

Currently there are two implementations of the \spo{} algorithm, described by Equation \ref{spo2equation}. The first is a basic implementation that is supplied by the manufacturer for testing purposes. This algorithm simply calculates the ratio of ratios without any filtering of data. We use this algorithm as a baseline for comparing other algorithms and our own implementation.

The second algorithm is an enhancement of the baseline algorithm that implements the same calculation of \spo{}, however it attempts to correct for the fact that the signal measured by the photodetector can be noisy. After performing baseline levelling of each signal, a Pearson correlation is calculated between the incoming red and infrared channels. Because the measurement site is the same the intensity of red and infrared light should be highly correlated. If they aren't correlated, it is most likely due to unwanted noise artifacts. Therefore, any signals that produce a correlation value below 0.4 are discarded. The code and a description of the algorithm is available through \footnote{https://www.instructables.com/id/Pulse-Oximeter-With-Much-Improved-Precision/}. 

Both aforementioned algorithms were originally implemented for an Arduino. For our analysis, we remove the Arduino relevant code and compile the remaining C code to allow for offline analysis and direct comparison between the two algorithms.

\subsection{Transmissive and Reflective}

\begin{figure}
\centering
\includegraphics[width=0.5\linewidth]{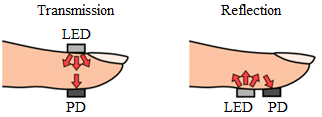}
\caption{Photodetector and LED placement for transmissive and reflective pulse oximeters (from \cite{tamura2014wearable}).}
\label{fig:oximetry_device}
\end{figure}

Pulse oximeters are available in two types; \textit{transmissive} and \textit{reflective}. These are characterized by the relative location of the LEDs and photodetector as shown in Figure \ref{fig:oximetry_device}.  

In \textit{transmissive} sensors the LEDs and photodetector sit across from each other so that when clipped to a user's finger, the light from the LED shines through the finger and into the photodetector. Medical settings, such as hospitals and clinics tend to employ transmissive sensors because they are more accurate. However, because they require the sensor to be clipped to the finger, can impede the wearers use of their hands, become uncomfortable after a few minutes, and are generally not well suited for continuous monitoring where the user requires mobility. Other points of attachment, such as the earlobe, are possible, but comfort and mobility remain an issue.


In \emph{reflective} pulse oximetry~\cite{mendelson1988noninvasive}, the photodetector sits beside the LEDs and measures light reflected off the user's tissue.  This allows measurement on a wider range of sites on the body (for example, the forehead) and provides greater mobility.
The downside to reflective sensors is that the overall amount of light received by the photodetector is less than in transmissive sensors, which means obtaining reliable data from them is more challenging~\cite{mendelson1988noninvasive}. Under ideal conditions, reflective sensors can still reliable enough to be used in hospitals, but generally only when comfort is more important, such as in NICUs \cite{Chen:2010:NBO:1870926.1871296}. In a mobile, wrist-worn device, however, conditions are far from ideal. Factors such as ambient light and motion can significantly degrade the quality of PPG data. While this is true for both reflective and transmissive sensors, because reflective sensors are already receiving a weaker signal, these factors have a much greater effect on reflective sensors.

Today reflective pulse oximeters are widely available in wrist-worn devices such as smart watches, fitness bands, and cell phones, but given the low accuracy of \spo{} measurement they are generally restricted to measuring heart rate.
In fact, major consumer devices have switched to using a single green LED for measurements as it provides better accuracy for heart rate, despite the green LEDs inability to measure \spo{}.

\subsection{Reflective Oximetry from Fingertip}




\begin{figure}[t]
    \includegraphics[width=\linewidth]{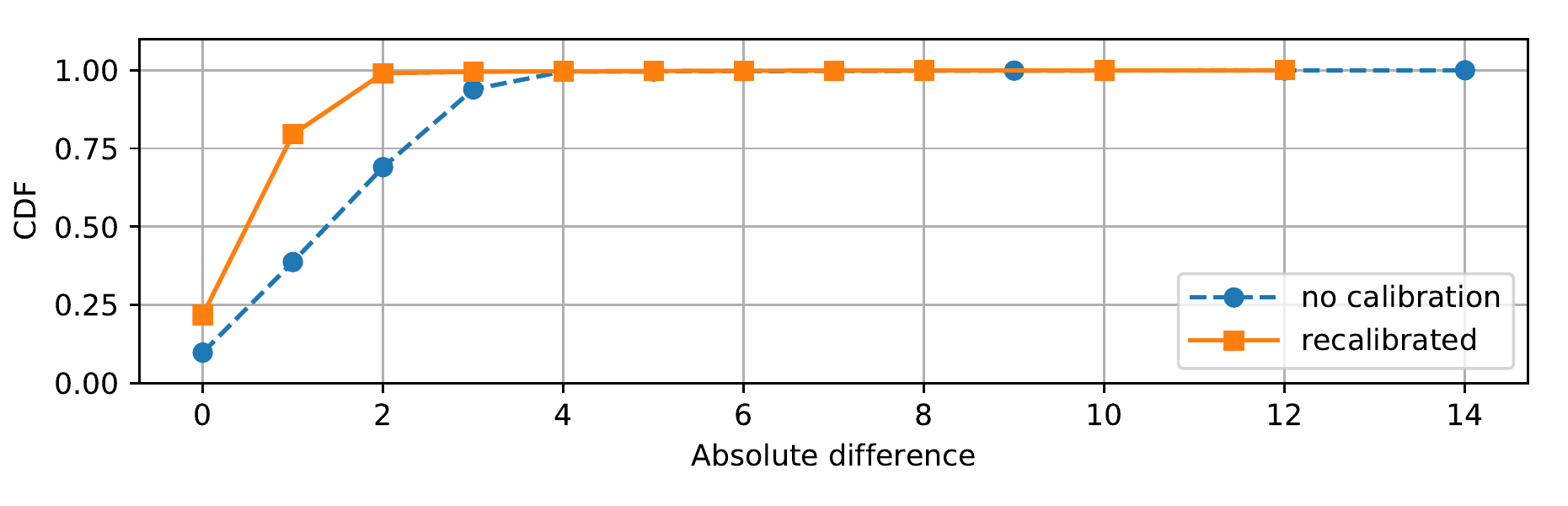}
    \caption{CDF of absolute difference between reflective and transmissive fingerprint sensor readings, before and after recalibration.}
    \label{fig:bias-calibration}
\end{figure}

We first show that a reflective $SpO_2$ sensor produces reliable measurements when placed on the fingertip.  For this purpose, we compare $SpO_2$ measurements collected with a $SpO_2$ reader we constructed using a MAX30102 reflective sensor (Section~\ref{max30102}), and measurements collected with a Berry BM3000B oximeter~\footnote{\url{http://www.shberrymed.com/usb-pulse-meter-bm3000b-p00037p1.html}}, a commercial device that uses a transmissive sensor.


%


We collected data from 10 subjects who wore the two $SpO_2$ readers on the non-dominant hand (reflective on the index finger, transmissive on the middle finger) for a period of 12 minutes each. 

The mean measurement difference between devices is $1.84\%$, with standard deviation of $1.32\%$ -- indicating good agreement between sensors except bias, which remains constant across all users. This bias is most likely a calibration error of one device or the other (the $y_0$ parameter).

We recalibrate the reflective sensor using readings in the first half of each measurement session, resulting in offset of $1.46$\% over the first half of the data. 
Figure~\ref{fig:bias-calibration} shows the difference between the recalibrated reflective and transmissive sensors over the second half of the data. After recalibrating the reflective sensor to remove bias, the mean absolute difference drops to $1.01\%$ with standard deviation of $0.77\%$ indicating strong agreement. Over 99\% of reflective sensor readings are within $\pm 2$\% of the transmissive.  Therefore, we conclude that our reflective $SpO_2$ sensor produces reliable measurements when placed on the fingertip.   In the reminder of this paper, we use  measurements collected with our reflective $SpO_2$ sensor mounted on the fingertip as ground truth.


\subsection{Reflective Oximetry from Wrist}

Unlike reflective fingerprint oximetry which is accurate, a na\"{i}vly applying existing methods to PPG traces obtained from the wrist results in unreliable $SpO_2$ measurements.

\begin{figure}[ht]
\centering
\begin{minipage}{0.5\textwidth}
    \includegraphics[width=\linewidth]{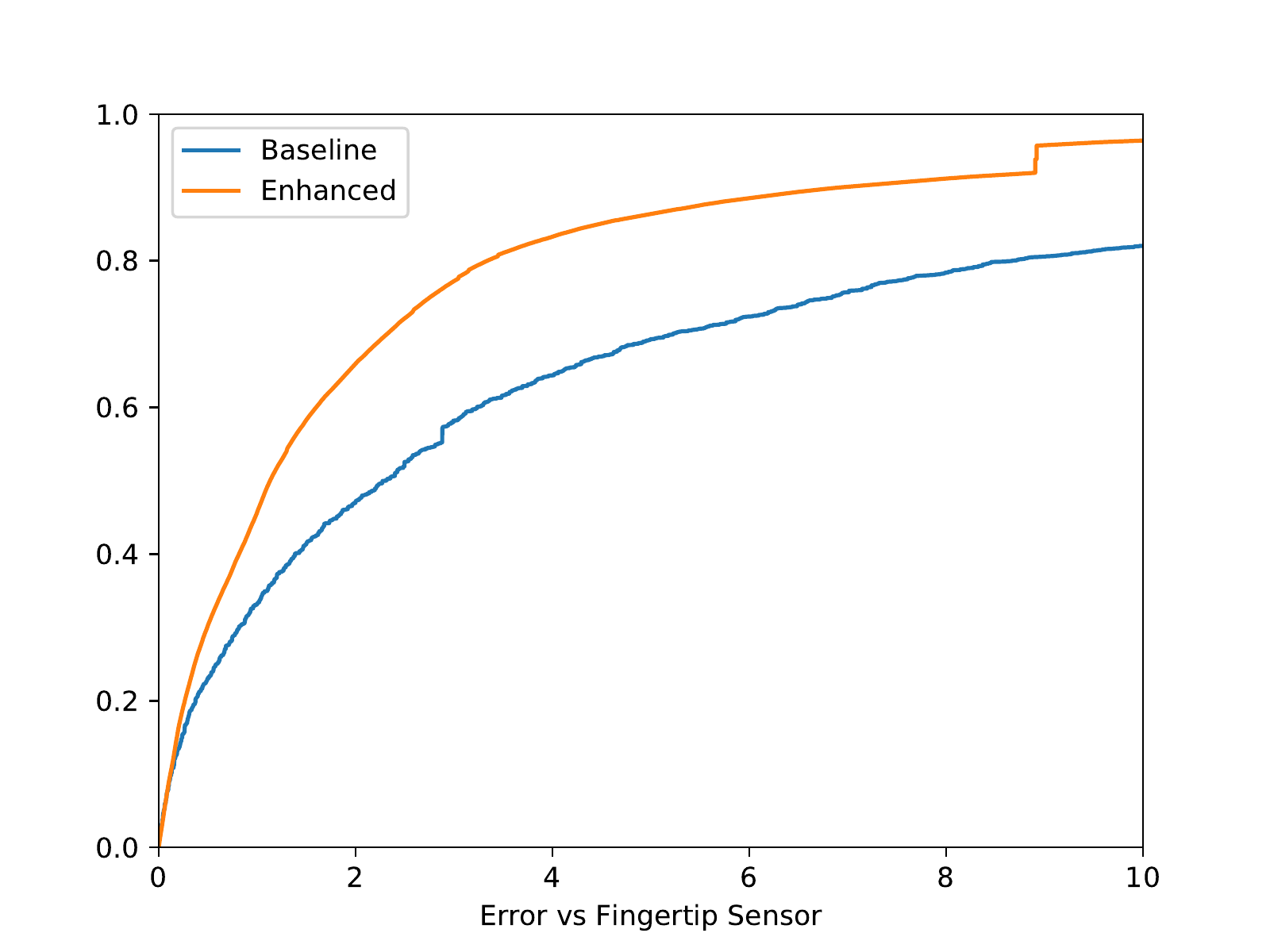}
    \caption{CDF of absolute difference between wrist and fingertip readings.}
    \label{fig:cdf_error}
\end{minipage}
\hfill
\begin{minipage}{0.46\textwidth}
    \includegraphics[width=\linewidth]{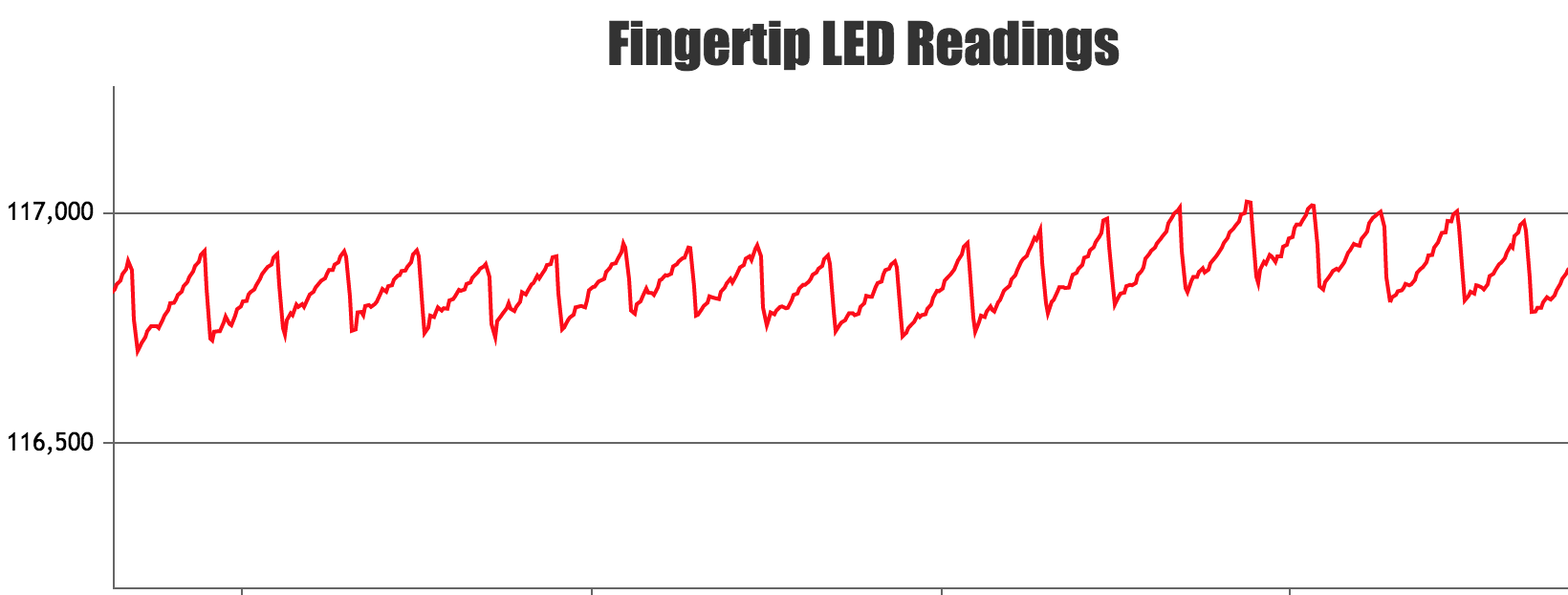}
    \includegraphics[width=\linewidth]{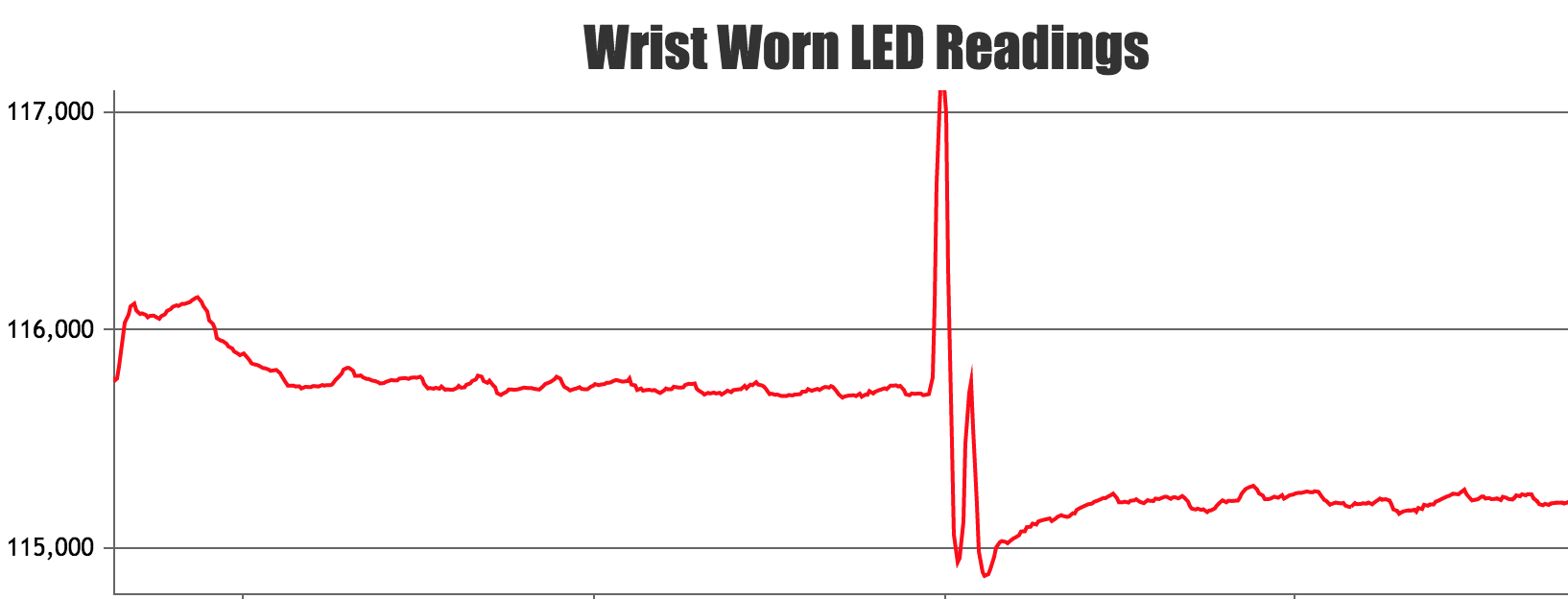}
    \caption{PPG trace for a fingertip vs. wrist attached sensor (taken from the PPG web platform described in Section~\ref{implementation}).}
    \label{ppgtrace}
\end{minipage}
\end{figure}

Figure \ref{fig:cdf_error} shows the CDF of absolute error of readings taken from the wrist using both existing algorithms, as compared to an identical fingertip sensor (Section~\ref{implementation} details our  on implementation and data collection).
Despite the increase in performance of the enhanced algorithm, more than 10\% of the readings across all users have an error of 5 percentage points or more compared to the fingertip readings, which we consider to be too big given that the healthy range for individuals is 90\% to 100\%.

Figure~\ref{ppgtrace} shows two PPG traces obtained from the same user at the same time using identical reflective sensors. The left PPG was captured from a fingertip-worn reflective pulse oximeter over several seconds. 
The strong periodic signal captures the change in flow of oxygenated blood through the fingertip. 
The right PPG was taken from the wrist. 
Even with a clean contact with the skin, this PPG is much noisier. 
The spike in the middle is caused by either motion or ambient light artifact, demonstrating how poor contact with the skin or a user's movements can cause errors and discontinuity in the signal. 
Algorithms used to produce reliable \({SpO_{2}}\) readings from a wrist-worn sensor must be able to mitigate and compensate for these errors.


\section{Approach}\label{approach}

This section outlines \codename, a new approach designed to identify which signal windows captured from the wrist-worn sensor will produce highly reliable \spo{} readings. 
We still employ the original algorithms for calculating \spo{}, however, the goal is to only apply this algorithm to signal windows that will produce a reliable reading.

To identify which signal window will produce a reliable reading, we employ statistical machine learning techniques to train a binary classifier that will classify input data as reliable or unreliable. As input to our classifier, we compute approximately 1000 features Tsfresh\cite{christ2018time} from 4 signal sources: red and infrared LEDs, gyroscope magnitude, and accelerometer magnitude. 
We use a signal window size of 100 sensor readings, or approximately 4 seconds of data when extracting features. This corresponds to the size of the window used to calculate \spo{} by the baseline algorithm. Intuitively, the window size used to calculate the \spo{} will have the greatest effect on it's outcome. To verify this assumption we explore other potential window sizes in section \ref{evaluation}.

We will use a level of agreement with a more reliably collected signal as our ground truth. Specifically agreement between the same sensor applied to both the wrist and fingertip. By taking the fingertip readings as truth, we mark wrist-worn readings as reliable if they are within a range of the fingertip readings. 

We use various thresholds of agreement between the wrist-worn and fingertip reflective sensors to create the reliability label for classification. Initially for experimentation we set this threshold to \(\pm 2.0\%\). That is, if the \spo{} output of the wrist-worn device is within 2 percentage points of the fingertip sensor, we mark the output of the wrist-worn device as reliable.  Although we use this threshold in a majority of experiments, we explore the classification results of other reliability threshold values in section \ref{evaluation}.

We score the classifier on precision, the ratio of true positive labels over the number of positive instances returned by the classifier, or:

\[ Precision\quad =\quad \frac { tp }{ tp+fp } \]

The precision of a reliability classifier is the ability of the classifier to only return with a positive score on a reliable result, and minimize the number of false positives. Although this will not produce reliable readings as frequently, it is more desirable for an \spo{} measurement device to provide few intermittent reliable results, rather than a continuous stream of potentially false readings. Intuitively, due to the relatively low fluctuations of true oxygen saturation measurements, \spo{} levels can be reliably interpolated with frequent enough measures. Therefore, we prefer a high true positive score, and a low false positive, with little concern for false negatives.

Considering precision allows us to quantify success of our classifier, we are ultimately concerned with reducing the error in calculated \spo{} readings. Therefore, for a second metric we use the room mean squared error (RMSE) of readings taken from the wrist-worn sensor as compared with the fingertip sensor. We take the RMSE before pruning values with \codename, and then calculate the RMSE after pruning to determine any improvement.

Because we can potentially remove a bulk of readings while pruning, we add a final metric described as the \textit{time between valid readings}. This measure describes the longest window of silence where \codename produces no reliable signal with which to calculate \spo{}. Although it is desirable for \codename to reduce our RMSE to zero, we do not want to prune signals so aggressively that we are left with readings that are too infrequent.

Section \ref{implementation} will describe the experimental setup used to accomplish these requirements, and section \ref{evaluation} will quantify the results of our trained classifiers.

\section{Implementation}\label{implementation}

Our work includes a hardware platform for collecting sensor data and a software platform for analyzing that data.

\subsection{Hardware}


Although it would have been desirable to utilize an existing consumer grade device to analyze the current state of wrist-worn pulse oximeters, we encountered two major issues when attempting to select one. The first issue is that the unreliability of \spo{} measurements taken from a wrist-worn pulse oximeter have led manufacturers to focus the technology solely on measuring a users heart rate. Most manufacturers only install a single LED, since heart-rate measurement algorithms implement peak detection and only rely on a single PPG trace. The second issue is that manufacturer APIs are too limited, and do not provide LED reflectance level needed for the PPG trace. In the devices we analyzed that did contain both LEDs required, such as the Apple Watch or various FitBits, the API access was limited to high level interpretations of biometric data from the user. Metrics like sleep quality, step counts, or heart rate were provided but access to the low level data was not. In order to adequately analyze the quality of PPG traces being received from a wrist-worn pulse oximeter, we we built our own wrist-worn device to measure SpO2. The sensors used and devices created for the purposes of experiments are described in the remainder of this section.

\begin{figure}[h]
\centering
\includegraphics[width=0.5\linewidth]{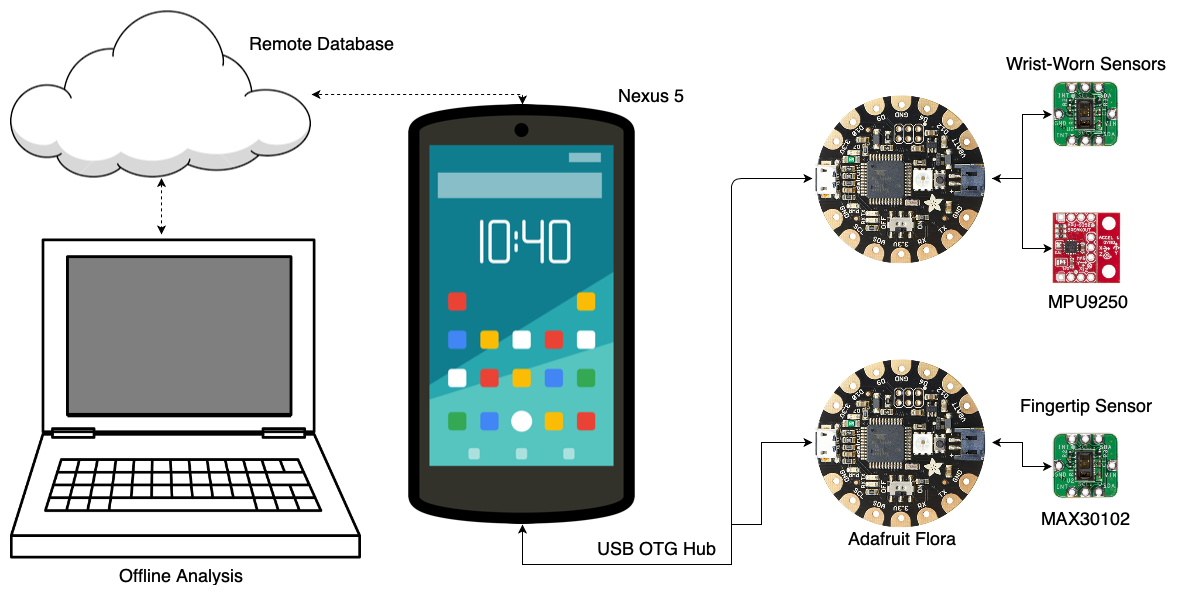}
\caption{The data collection platform and sensors.}
\end{figure}

\subsubsection{MAX30102 Sensor}
\label{max30102}

We use the MAX30102~\cite{max30102} reflective pulse oximeter from Maxim Integrated for our data collection. The sensor's provided by Maxim Integrated are the same as those commonly used by manufacturers such as Samsung for oxygen saturation measurement on smart phones and other devices. The sensor is described by the manufacturer as \textit{an integrated pulse oximetry and heart-rate monitor biosensor module}. It provides red and infrared source LED's onboard the chip with an adjacent photodetector. Communication with microcontrollers is accomplished via the \(I_{2}C\) protocol and where the sensors publish readings with a sample rate of 25Hz.

\subsubsection{Wearable Prototype}

In order to measure \spo{} from a user with confidence, sensors are used to take measurements from two points of contact on a single user, namely the fingertip and the wrist. During experiments, PPG traces and other data are collected from all sensors simultaneously. 

\subsubsection{Wrist-Worn Reflective Sensor with Motion Tracking}\label{wrist-worn-device}

This is the primary sensor platform. The wearable consists of two sensors, including the MAX30102 sensor described in section \ref{max30102}, and an MPU9250 IMU sensor to track acceleration and rotation of the wrist worn device. Readings from the two sensors are captured and aligned using an Adafruit FLORA microcontroller. The three components are sewn into a fitness band for stability and consistency across measurements. The wrist-worn device is attached to the dominant hand of a user during experiments. The device allows for users to maintain range of motions in their wrist and movement throughout the duration of experiments is encouraged. The implications of using the methods described in this paper on a custom device versus a consumer grade device are discussed in section \ref{future}. The user wears the device with the pulse oximeter facing the top of the wrist so that it matches the sensor placement in a vast majority of consumer grade wristbands and smartwatches.

\begin{figure}[!h]
\centering
\begin{minipage}{.56\linewidth}
\includegraphics[width=1.0\linewidth]{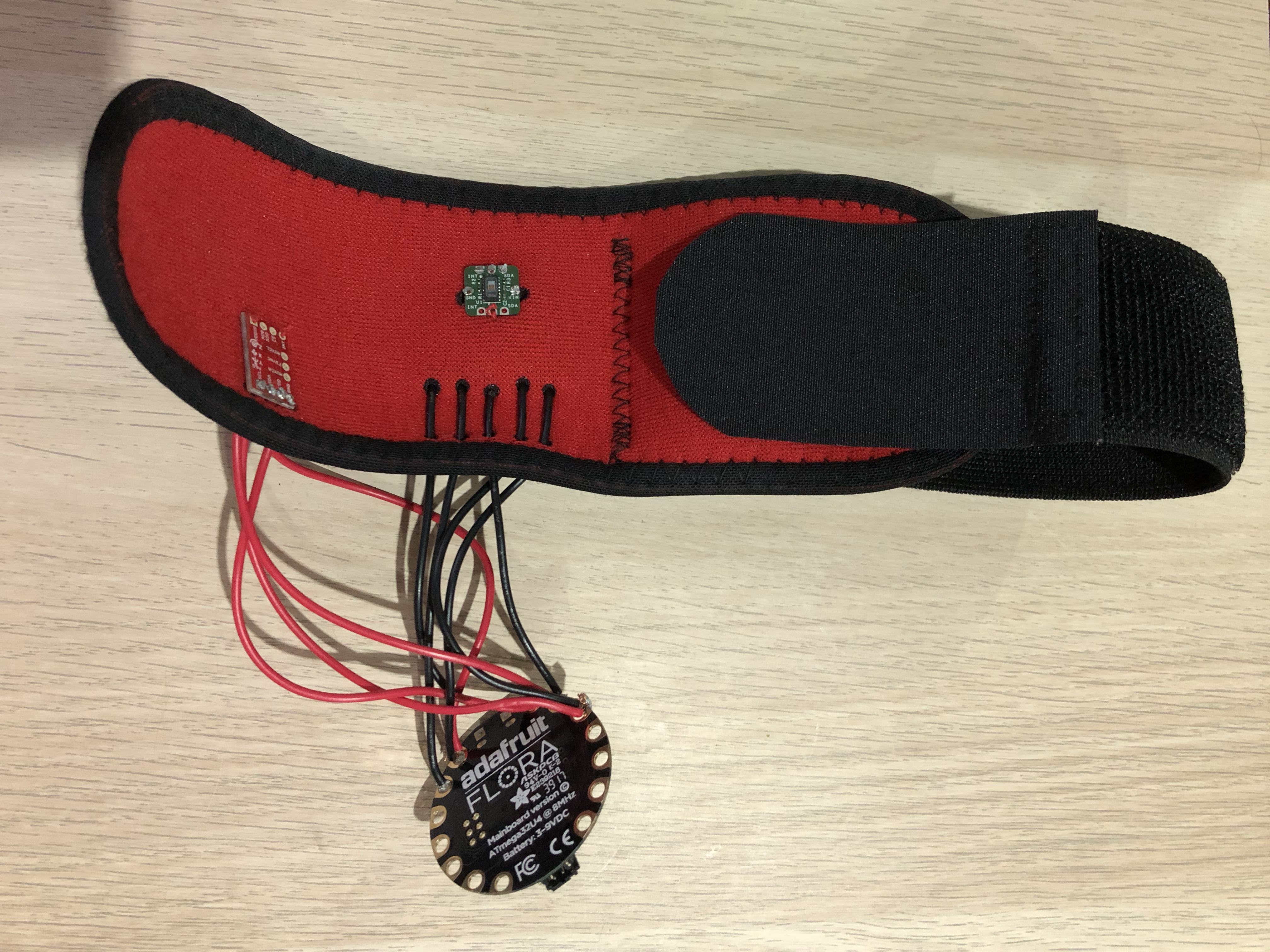}
\end{minipage}%
\hfill
\begin{minipage}{.42\linewidth}
\includegraphics[width=1.0\linewidth]{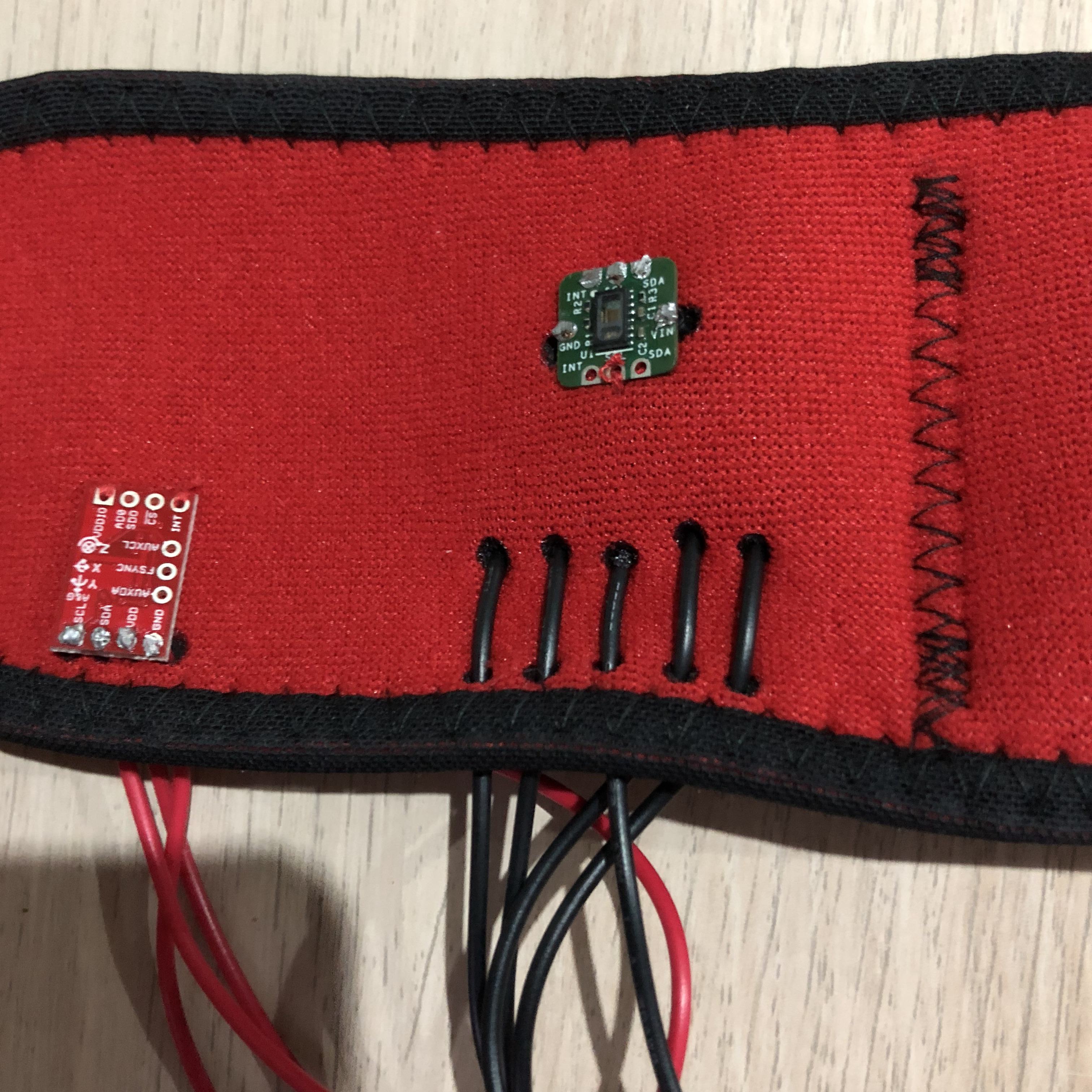}
\end{minipage}%
\caption{Custom Wrist Wearable and Sensor Bed}
\end{figure}

\subsubsection{Fingertip-Worn Reflective Sensor}

To establish a baseline for best-case signal from the MAX30102 sensor, we attach a second sensor to the index fingertip of the non-dominant hand of the user. The sensor is attached with medical tape to ensure a consistently applied pressure. The signal is again captured using an Adafruit FLORA microcontroller.

Our ground truth uses the exact same sensor applied to both the wrist and fingertip.
This eliminates a variable in the experiment: we are interested in reliability across different measurement sites (specifically, the wrist vs fingertips), rather than across different hardware manufacturers. The MAX30102 was demonstrated to be a reliable sensor for measuring signals from the fingertip in section \ref{background}.

\subsubsection{Collecting and Aligning Sensor Signals}

We wrote a custom Android application to capture and visualize signals from all sensors. The Android application communicates with each device over the USB serial protocol. A USB hub is used to communicate with the devices simultaneously as well as to provide power to each device. To later align the readings between devices, a timestamp is attached by the Android application when each reading is received. Finally, the application saves the collected readings to a remote database for offline processing.

\subsection{Software}

In addition to the Android application used to visualize data streams, we developed several Python applications to clean, align, and transform incoming data to be used with various out of the box machine learning libraries.


\subsubsection{Reliability Classifier}

To train our classifier, we first extract features from the wrist-worn sensors to be used as inputs when predicting the reliability of the signal. We use two radiance signals from the red and infrared LEDs, and two signals from the magnitude of the gyroscope and accelerometer in the MPU9250. As discussed in section \ref{related}, it has been shown that motion of the device can be used to detect noise in the PPG reading. The magnitude of various motion readings from the IMU are used to automatically filter motion artifacts during classification.

We use the Tsfresh~\cite{christ2018time} Python library to extract features from our time series data. The library automatically selects features by calculating a comprehensive set of features on the provided data and then pruning the list of features based on the provided labels being predicted. Feature significance is performed using the Benjamini Hochberg procedure \cite{benjamini2001control}. Depending on the training data provided, approximately 900-1000 features are selected by the library. Many features have a very low significance and can be pruned without affecting classifier performance. 
Table \ref{top-features} shows the top 15 features of the 1000 total features. The first argument is the channel used, and additional parameters are dependent on the feature extracted. Detailed descriptions of the features and the Python API used to generate them are available at \footnote{https://tsfresh.readthedocs.io/en/latest/text/list\_of\_features.html}.  Based on these tests, the infrared LED channel adds the most information with a smaller dependency on the red LED and gyroscopic magnitude channels.

\begin{figure}[h]
\begin{minipage}[t]{0.48\textwidth}
    \centering
    \captionof{table}{Top 15 features.}
    \label{top-features}
    \begin{tabular}{ l }
    \toprule
    \textbf{Tsfresh Function Call}                           \\ 
    \midrule
    longest\_strike\_below\_mean(`ir')                       \\ 
    autocorrelation(`ir', 6)                                 \\ 
    autocorrelation(`ir', 5)                                 \\ 
    autocorrelation(`ir', 7)                                 \\ 
    autocorrelation(`ir', 8)                                 \\ 
    autocorrelation(`ir', 9)                                 \\ 
    cid\_ce('ir', normalize=True)                            \\ 
    autocorrelation(`ir', 4)                                 \\ 
    ar\_coefficient(`red', \{"coeff": 0, "k": 10\})          \\ 
    spkt\_welch\_density('red', {“coeff”: 2})                       \\ 
    ar\_coefficient(`ir', \{"coeff": 0, "k": 10\})           \\
    mean(`gyro')                                             \\
    sum\_values(`gyro')                                      \\
    fft\_coefficient(`gyro', \{"coeff": 0, "attr": "abs"\})  \\
    fft\_coefficient(`gyro', \{"coeff": 0, "attr": "real"\}) \\ 
    \bottomrule
    \end{tabular}
\end{minipage}
\hfill
\begin{minipage}[t]{0.48\textwidth}
    \centering
    \captionof{table}{Optimal XGBoost parameters.}
    \label{table:parameters}
    \begin{tabular}{ l l }
    \toprule
    \textbf{Parameter}   & \textbf{Value}  \\
    \midrule
    Learning Rate        & 0.1             \\ 
    Estimators           & 100             \\ 
    Max Depth            & 3               \\ 
    Minimum Child Weight & 3               \\ 
    Regularization Alpha & 0.3             \\ 
    Subsample Ratio      & 0.9             \\ 
    Objective            & Logistic Binary \\ 
    \bottomrule
    \end{tabular}
\end{minipage}
\end{figure}

To select the classifier that best generalizes to unseen data, we compare several binary classifiers available in the scikit-learn library~\cite{scikit-learn} using multiple validation sets. We found gradient boosting classifiers to provide robustness, generalizability, and the most consistently usable results. We use the XGBoost library \cite{Chen:2016:XST:2939672.2939785}, as it provides similar results with faster training times. We perform a cross validated grid search of hyperparameters to further tune the classifier. Evaluating performance based on data trained across several individuals yields the optimized parameters shown in Table~\ref{table:parameters}. Data was trained using 10-fold cross validation, individual folds were checked against multiple separate validation sets.

During training in all experiments, non-overlapping windows are used to ensure that feature data is independent. Not only does preventing overlap ensure data remains i.i.d., but it also reduces feature extraction and training time by a factor equal to the windowing size, which is 100 in a majority of experiments. When applying the trained classifiers to unseen data, sliding windows of features with a step-size of 1 are taken. This ensures that we have the maximum number of output results with which to analyze.

\subsection{Data Collection}

We collect data from 10 participants. Each user has the wrist-band with the pulse oximeter and IMU sensor attached to their dominant hand, and a MAX30102 sensor attached directly to their fingertip on the opposing hand. Trials on each participant are conducted for approximately 12 minutes, during which time users are encouraged to continue using their dominant hand in an effort to provide the most naturally acquired readings. To reduce motion artifacts when acquiring ground truth readings, participants are asked to keep their non-dominant hand motionless for the duration of the experiment. Table \ref{table:users} shows a summary of the 10 participants. Users range from 20-55 years of age and vary greatly in skin colour. 18000 readings are used from each user, corresponding to 12 minutes of readings acquired at a rate of 25Hz.

\begin{table}[ht]
\centering
\caption{Participant Information and proportion of reliable labels for each.}
\label{table:users}
\begin{tabular}{ c c l c }
\toprule
\textbf{User} & \textbf{Age} & \textbf{Relative Skin Tone} & \textbf{Proportion of reliable readings} \\
& & & \textbf{(within $\pm2$ of fingertip)} \\
\midrule
1             & 28           & Light        & 47.3\% \\
2             & 24           & Light        & 17.6\% \\
3             & 32           & Dark         & 72.0\% \\
4             & 38           & Light        & 44.8\% \\
5             & 20           & Dark         & 0.3\%  \\
6             & 31           & Medium       & 28.3\% \\
7             & 24           & Dark         & 2.1\%  \\
8             & 31           & Medium       & 6.0\%  \\
9             & 26           & Light        & 7.3\%  \\
10            & 55           & Light        & 16.5\% \\
\bottomrule
\end{tabular}
\end{table}

We also how how much of the data collected from the wrist sensor is within 2 percentage points of the fingertip sensor. This shows that without adequate filtering, generally, very few readings from the wrist are accurate. Also notable, is the drastic variation between users in the proportion of readings considered reliable. These differences could stem from a large variety of variables that cannot be controlled in the wild. Variables such as; skin colour, device tightness, wrist thickness, movement, and ambient light, can all affect how much of a signal collected from a user is reliable.

\section{Experimental Evaluation}\label{evaluation}

We evaluate the ability of \codename to correctly identify clean PPG readings for the purposes of measuring peripheral oxygen saturation. Through our data analysis we answer the following questions:

\begin{enumerate}
\item How does \codename perform compared with existing algorithms?
\item How effective is \codename across different measurement sites?
\item Does classification translate to unseen skin tone colours?
\item Can we calibrate \codename on a per-user basis?
\item How does the IMU effect performance?
\item How do domain-specific hyper-parameters effect performance?
\end{enumerate}

Our main performance metric is the root mean square error (RMSE) of readings taken from the wrist, compared to the fingertip ground truth.
We also evaluate the precision of the \codename classifier, as defined in Section~\ref{approach}.

\begin{figure}[ht]
  \centering
  \begin{minipage}{0.48\linewidth}
    \includegraphics[width=1.0\linewidth]{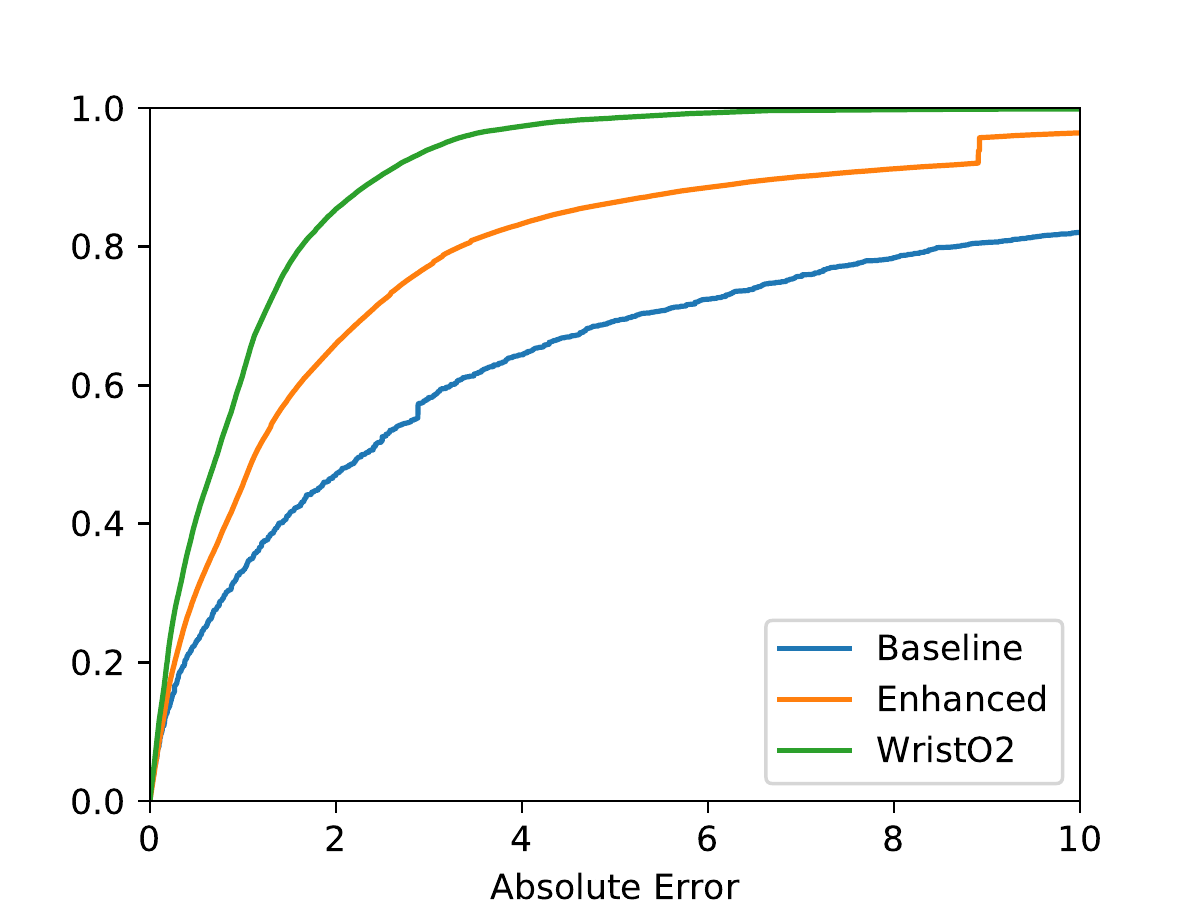}
    \caption{CDF of RMSE for existing algorithms and \codename for all users.}
    \label{fig:cdf-error}    
  \end{minipage}
  \hfill
  \begin{minipage}{0.48\linewidth}
        \includegraphics[width=1.0\linewidth]{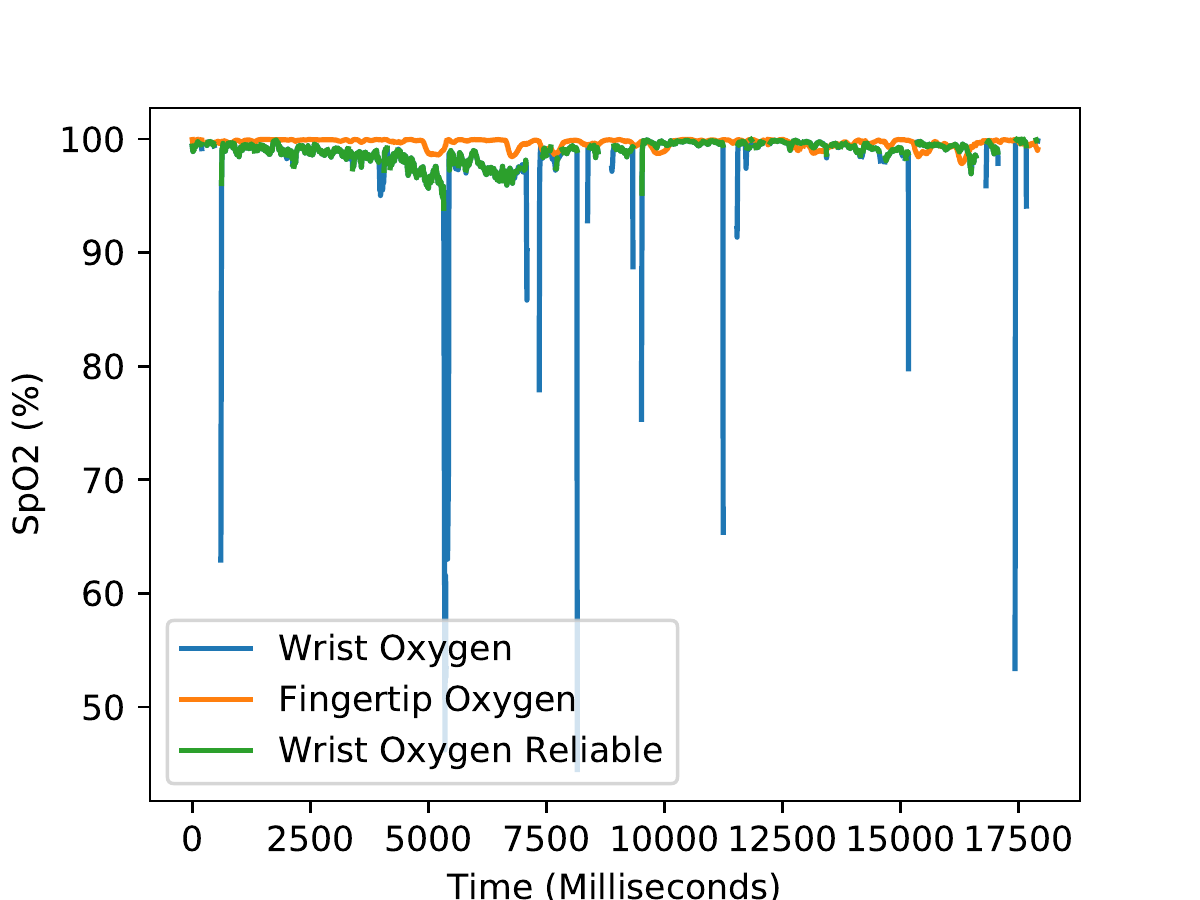}
    \caption{Trace of existing algorithms and \codename for a single user.}
    \label{fig:classifier-vis}    
  \end{minipage}
\end{figure}

\begin{table}[ht]
\centering
\caption{Mean (Std. Dev.) of classification results across all 10 users, trained and tested on the top of the wrist.}
\label{table:performance-all}
\begin{tabular}{@{}cccc@{}}
\toprule
\textbf{\codename Precision} & \textbf{Baseline RMSE} & \textbf{Enhanced RMSE} & \textbf{\codename RMSE} \\ \midrule
73\% (19\%)        & 14.5\% (6.9\%)          & 6.7\% (4.4\%)          & 1.5\% (0.7\%)           \\ \bottomrule
\end{tabular}
\end{table}

\subsection{Performance of \codename}
\label{overall}

We perform leave-one-out cross validation across participants, where a single participant's signal is classified given training data from all others.
Figure \ref{fig:cdf-error} shows the resulting absolute errors of wrist sensor readings across all readings for both existing algorithms and \codename, and Table \ref{table:performance-all} shows a summary of classification and algorithm performance.
Pruning results with \codename shows a drastic reduction in error compared to existing methods. 
\codename reduces RMSE of \spo{} measurement by an order of magnitude compared to the baseline algorithm, and by more than 4 times for the enhanced algorithm.

Figure \ref{fig:classifier-vis} shows a trace for a single user, to better illustrate the effect of \codename, 
The blue line representing the enhanced algorithm applied to a PPG trace collected from the wrist over 12 minutes, and the orange line representing the enhanced algorithm applied to the signal collected from the fingertip during the same session. Finally, the green line in Figure \ref{fig:classifier-vis} represents the readings remaining after \codename prunes unreliable results. 
Spikes and inaccuracies are clearly visible even with the Enhanced algorithm. \codename successfully rejects  many of these unreliable readings.


\begin{figure}[th]
  \centering
    \includegraphics[width=0.5\linewidth]{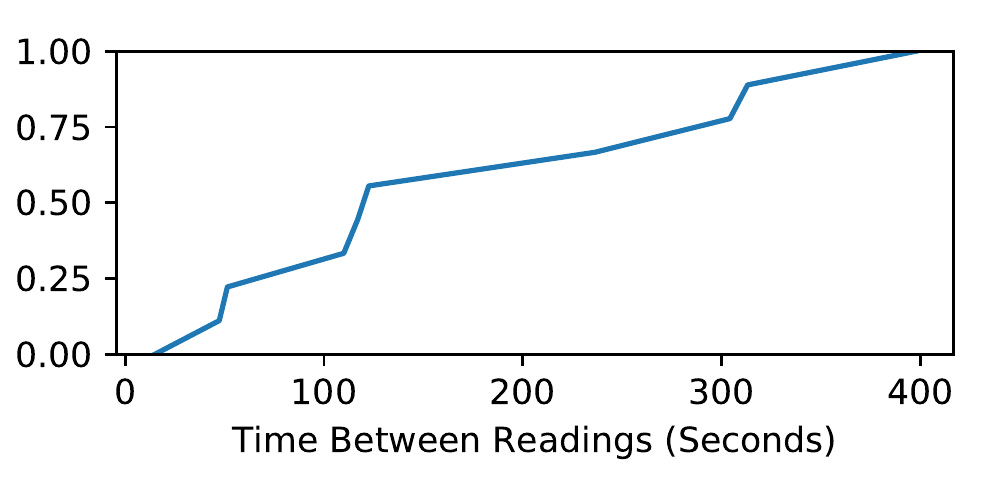}
    \caption{CDF of longest delay between valid readings.}
    \label{fig:cdf-time}
\end{figure}

The reduction in error comes at the cost of producing less readings compared to the existing algorithms. 
Figure \ref{fig:cdf-time} shows the CDF of the maximum size of an interval with no reliable values for all users across the leave-one-out validation. In other words, the longest silent window for each user where the classifier is returning no reliable readings.
The worst case scenario for a no reading window was approximately 6 minutes and 40 seconds, while the average worst-case across all users is approximately 3 minutes. Given that the method for acquiring reliable readings currently requires a user to actively clip a commercial pulse oximeter to their fingertip and wait, the time between readings from existing methods would be collected in the order of several hours or even half a day. A mean interval of less than 3 minutes for automatic collection of reliable readings is a dramatic improvement.

\subsection{Comparing Measurement Sites}

Each participant listed in table \ref{table:users} participated in a second trial where the wrist-worn pulse oximeter was applied to the bottom of the wrist, rather than the top. Using the same approach described in section \ref{overall}, we run leave-one-user-out cross-validation to test the performance of \codename with a signal collected from the bottom of the wrist. Table \ref{table:bottom} presents the results.

\begin{table}[ht]
\centering
\caption{Mean of classification results across all 10 users for varied measurement sites.}
\label{table:bottom}
\begin{tabular}{@{}llllll@{}}
\toprule
\textbf{Training Site} & \textbf{Testing Site} & \textbf{\codename} & \textbf{Enhanced} & \textbf{\codename} & \textbf{Participants with} \\ 
  &   & \textbf{Precision} & \textbf{RMSE} & \textbf{RMSE} & \textbf{no readings} \\ 

\midrule
Bottom of Wrist        & Bottom of Wrist       & 23\% (33\%)        & 14.7\% (13.1\%)      & 12.0\% (3.7\%)      & 5       \\
Top of Wrist           & Bottom of Wrist       & 25\% (28\%)        & 14.7\% (13.1\%)      & 8.9\% (10.3\%)      & 2       \\ \bottomrule
\end{tabular}
\end{table}

The 10 users had an average RMSE of 14.68\% when measurements were taken from the bottom of the wrist with the enhanced algorithm. After pruning using the classifier trained on this data, the average RMSE is reduced to 12.0\%. Despite the slight improvement, it should be noted that for half of the users, \codename pruned all values, meaning no readings in the 12 minute window were marked as reliable.

Applying \codename to the bottom of the wrist when trained on signals from the top shows improved. The original 14.68\% RMSE is further reduced to 8.84\%. Furthermore, only two out of the ten trials in this instance provided no reliable readings. So, given a situation where data must be collected from the bottom of the wrist, the original classifier trained in section \ref{overall} can still be used to prune less reliable labels. It is also notably less aggressive in pruning than the classifier trained on readings from the bottom of the wrist.

It is clear that it better to collect traces, and train classifiers using data acquired from the top of the users wrist. This is a positive result when considering that this is where almost all consumer grade devices choose to collect data from already.

\subsection{Effect of Skin Tone}

As discussed in section \ref{related}, it has been shown that it is more difficult to collect a reliable signal when darker pigment exists on the skin, whether naturally or artificially from tattoo ink. This section aims to quantify potential difficulty in collecting reliable PPG traces from users of various skin tones. Five of the participants had skin colour that we qualitatively define as light, relative to the other users. We separate the users qualitatively into two groups, lighter- or darker-skinned, and train classifiers with all permutations of these groups. Mean (and Std. Dev.) are shown across users of the \textit{Testing Group}. In cases where the training and testing groups are the same, leave-one-out cross-validation is used across user's of the group.

\begin{table}[ht]
\centering
\caption{Effects of skin tone with various training and testing permutations.}
\label{table:skin-tone}
\begin{tabular}{@{}llccc@{}}
\toprule
\textbf{Training Group} & \textbf{Testing Group} & \textbf{\codename Precision} & \textbf{Enhanced RMSE} & \textbf{\codename RMSE} \\ \midrule
Dark                    & Dark                   & 37\% (28\%)        & 8.0\% (5.0\%)        & 1.6\% (0.5\%)       \\
Dark                    & Light                  & 80\% (20\%)        & 5.2\% (3.2\%)        & 1.3\% (1.0\%)       \\
Light                   & Dark                   & 42\% (32\%)        & 8.0\% (5.0\%)        & 4.3\% (3.3\%)      \\
Light                   & Light                  & 69\% (27\%)        & 5.2\% (3.2\%)        & 2.6\% (2.0\%)       \\ \bottomrule
\end{tabular}
\end{table}

Table \ref{table:skin-tone} shows that precision is improved when classifying on lighter skin as opposed to darker skin, regardless of the skin-tone used during training. Unexpectedly, results are slightly improved for predicting lighter skin signal reliability when the darker skinned group was used for training. Given the high variance in results, it is likely that this discrepancy is due to the small sample size. There is also slightly less data with the light-to-light experiment since leave-one-out cross validation is used. Utilizing 4 users for training, instead of 5 for the dark-light experiment, means less data is available for training and performance could be affected.

We caution that sample size is too small to draw strong conclusions about the magnitude of effects, and much more data will be needed to prove performance discrepancies between pigment groups. Regardless, in both groups the error is reduced by \codename; and we have shown that the classifier will generalize to pigment colours that it was not trained on.

\subsection{Per-User Training}

This experiment attempts to show the viability of \codename to provide on-the-fly training to build a personalized classifier on a per user basis. Consider a user that has a wrist-worn device with a pulse oximeter capable of measuring \({SpO_{2}}\), such as a smart watch, and a similar fingertip sensor such as those that exist in the back of various Samsung smart phones. During a calibration phase, the user can be instructed to wear the smart watch while simultaneously pressing their finger against the sensor on the smart phone. Once sufficient calibration data can be captured, aligned, and preprocessed, the classifier can be retrained with the additional data to provide the user with more reliable readings from the wrist-worn device.

\begin{table}[]
\centering
\caption{Adding user calibration data to increase \codename performance.}
\label{table:calibration}
\begin{tabular}{@{}lccc@{}}
\toprule
\textbf{Calibration Data} & \textbf{\codename Precision} & \textbf{Enhanced RMSE} & \textbf{\codename RMSE} \\ \midrule
None                      & 33\%               & 9.3\%                & 3.8\%               \\
+ 2 minutes               & 34\%               & 9.3\%                & 3.3\%               \\
+ 10 minutes              & 41\%               & 9.3\%                & 3.1\%               \\ \bottomrule
\end{tabular}
\end{table}

We train a classifier with 9 users and test it on an unseen user. We then retrain the classifier with 2 minutes of additional calibration data, and again with 10 minutes. The results are summarized in Table \ref{table:calibration}. Pruning signal windows with \codename reduces the RMSE to 3.8\% even when no calibration data is used. Using a small amount of user specific training data on top of the original training set further reduces the RMSE up to 0.7\%.

\subsection{Importance of Accelerometer and Gyroscope}

To study the effect of the features extracted from the IMU signal on classification, we run a similar experiment to section \ref{overall} with varied combinations of features from the LEDs and IMU sensor. Table \ref{table:motion} summarizes the results of the experiments

\begin{table}[h]
\centering
\caption{Effects of IMU features on classification.}
\label{table:motion}
\begin{tabular}{@{}lcccc@{}}
\toprule
\textbf{Signal Channels} & \textbf{Num. Features} & \textbf{\codename Precision} & \textbf{Enhanced RMSE} & \textbf{\codename RMSE} \\ \midrule
LED Only                 & 489                    & 69\% (19\%)        & 6.7\% (4.4\%)        & 1.8\% (0.9\%)       \\
IMU Only                 & 497                    & 47\% (35\%)        & 6.7\% (4.4\%)        & 5.5\% (5.2\%)       \\
LED + IMU                & 986                    & 73\% (19\%)        & 6.7\% (4.4\%)        & 1.5\% (0.7\%)       \\ \bottomrule
\end{tabular}
\end{table}

Approximately half of the features extracted and selected by the Tsfresh pipeline are features from the IMU. Although features from the LED channels alone contribute to a significant reduction in the RMSE, adding the 497 features extracted from the IMU signals further reduces the RMSE to a very low 1.5\%. It is sensible that the LED channels contribute a majority of the performance increase considering the LEDs are used directly to calculate $SpO_2$. We verify that this is the case by training the classifier with traces solely from the IMU, which shows a negligible increase in performance.

\subsection{Effects of varied thresholds and window sizes}

This section aims to tune two parameters discussed in section \ref{approach}, namely the window size for feature extraction, and the threshold of reliability.

\begin{figure}[ht]
  \centering
  \begin{minipage}{0.48\linewidth}
    \includegraphics[width=1.0\linewidth]{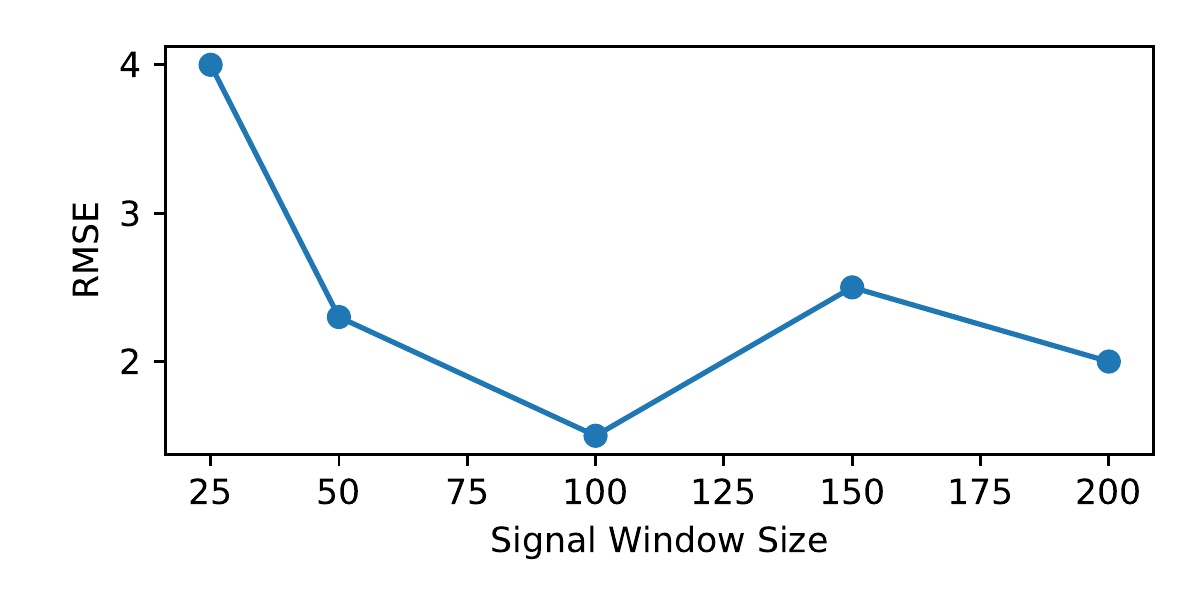}
    \caption{RMSE for different signal window size.}
    \label{fig:window}    
  \end{minipage}
  \hfill
  \begin{minipage}{0.48\linewidth}
        \includegraphics[width=1.0\linewidth]{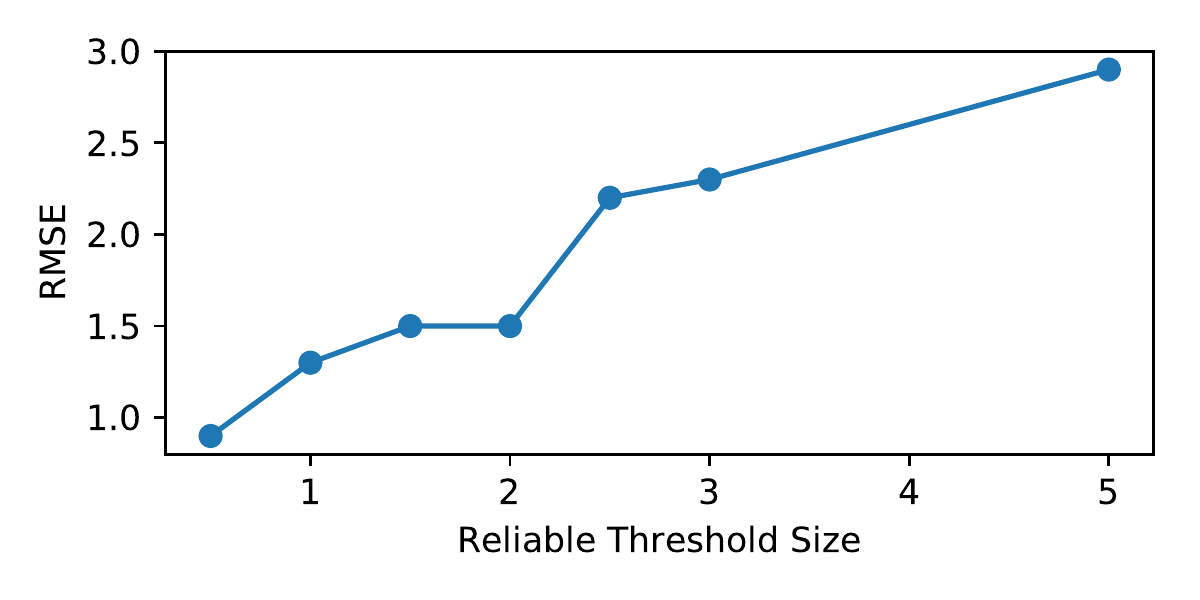}
    \caption{RMSE for different thresholds.}
    \label{fig:threshold}    
  \end{minipage}
\end{figure}

\subsubsection{Window Size}

Figure \ref{fig:window} shows varied window sizes and their resulting RMSE with a window size of 100 providing the lowest. However, it is useful to know that acceptable results can be achieved with smaller window sizes. This is useful if performance is an issue. Because the feature extraction takes longer with larger window sizes, we can use smaller feature windows at less frequent intervals to decrease feature extraction time with only a small performance penalty.

\subsubsection{Reliability Threshold}

\begin{figure}[ht]
\centering
 \includegraphics[width=0.5\linewidth]{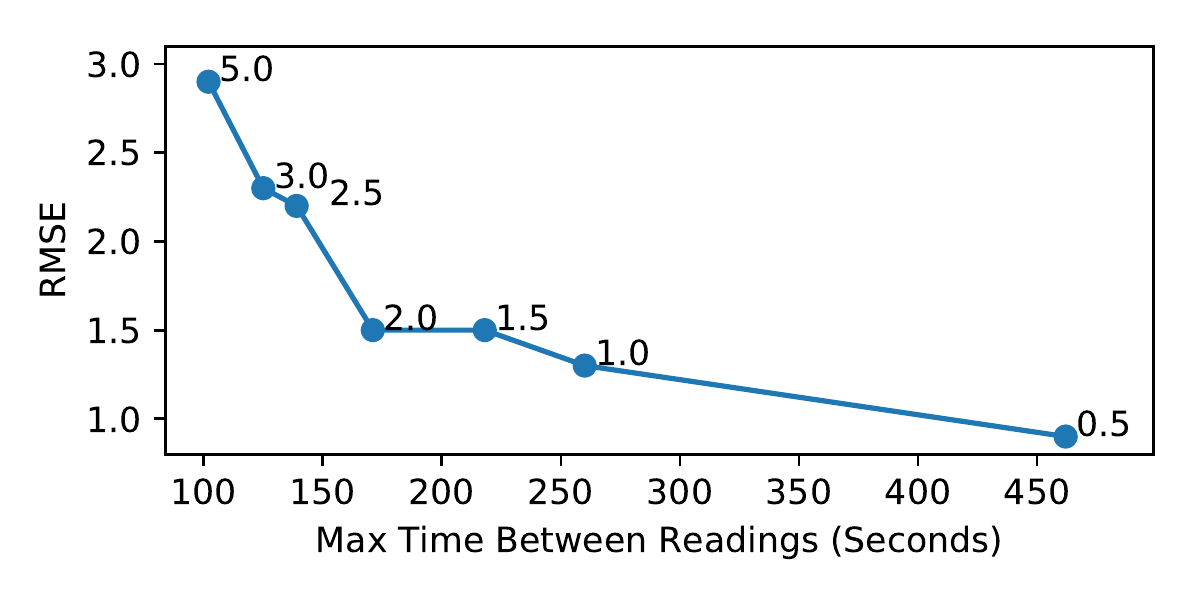}
\caption{The trade-off between quantity and quality of readings for different reliability thresholds.
For every threshold, the X axis shows the resulting worst-case interval between reliable readings, and the Y axis shows the resulting RMSE.}
\label{fig:threshold-time}    
\end{figure}

Figure \ref{fig:threshold} shows varied sizes of reliability thresholds and their corresponding RMSE results. Although it would appear that lower threshold values improve results overall, it is worth noting that the frequency of acquired readings is inversely proportional to the expected RMSE, as Figure \ref{fig:threshold-time} demonstrates for the varied threshold sizes.

\section{Discussion}
\label{future}

This section covers future work for \codename, including necessary steps for deploying it in the wild on consumer grade devices.

{\bf Reducing Compute Costs}  For \codename{} to be deployed on existing wrist-worn mobile devices, measures must be taken to ensure the performance and battery life of mobile devices are not affected. First, the feature set could be pruned far enough that the computational cost of feature extraction is reasonable to perform on live data directly onboard the device taking the measurements. In addition to pruning features, work similar to Sidewinder\cite{liaqat2016sidewinder} could be used to offload signal reliability calculation to a lower powered processor, and subsequently wake the device when a usable signal is detected.

Alternatively, instead of optimizing the feature extraction, we could utilize a cloud service to stream the collected data, offload the computation, and collect the results. The data is small enough that an hours worth of data can be transferred within seconds to a remote server that processes the data.

{\bf Improving Classification}  Although the classifier is trained, tested, and validated on a diverse group of people, the small number of people used in the study could be limiting the ability of the classifier to predict some values if they are all within a healthy range. Future work will include more participants, and incorporate participants with lower \spo{}. 

We are also considering extending the classifier to multi-label classification or regression. That is, predict not whether a signal will produce a reliable label within a certain threshold, but predict the confidence that the label will be produced within various different thresholds. For example 1\%, 3\%, and 10\%. We leave this to future work.

{\bf Extending to Existing Wearable Devices}  The next obvious iteration of the wrist-worn device is to either build or utilize a consumer grade device. \codename could be applied in consumer grade devices if low level access to the LED sensor were to be provided. The use of a consumer grade device could potentially improve results solely based on the quality of the hardware. 

\section{Related Work}\label{related}

To our knowledge this is the first work that applies state-of-the-art feature extraction and machine learning approaches to increase the reliability of \({SpO_2}\) measurements taken from the signals of wrist-worn devices.

Much of existing work on reflective sensors focused on heart rate measurement, such as rule based detection of heart rate for reliability \cite{al2004pulse}. Ra et al. perform reliability detection in the context of wrist heart-rate measurement using hidden Markov models applied to a single LED source on existing smart watches \cite{Ra:2017:IAS:3032970.3032977}. There has been work done to improve reading reliability in fingertip sensors at the algorithmic level for both heart rate and \({SpO_2}\) through signal preprocessing and noise reduction \cite{mohan2016measurement}\cite{yao2005short}. Possible wearability sites, including the wrist, and various sensor configurations have been considered in the context of telehealth monitoring \cite{mendelson2003measurement, mendelson2006wearable}. Other work has documented the process of building transitive pulse oximeters from scratch \cite{bagha2011real}. Reflective pulse oximeters are widely used and studied in medicine in places where transitive pulse oximeters are not feasible, such as infant monitoring \cite{tobler2001fetal}.

Accuracy and reliability of fingertip worn pulse oximeters have been analyzed in great detail, such as quantifying quality of \({SpO_2}\) measurements in patients with specific conditions or qualities. Severinghaus et al.\cite{severinghaus1990effect} showed that bias in \({SpO_2}\) measurements increases during a state of anemia (low red blood cell count). Emery et al. \cite{emery1987skin} and Cote et al. \cite{cote1988effect} showed the effects of dark skin pigmentation and ink in convoluting measurements of fingertip worn pulse oximeters. Additionally Lee et al. \cite{lee1993factors} showed that lower true pulse oximetery values were overestimated for a specific set of people from Singapore due to darker pigmentation. 

Yao et al.\cite{yao2004novel} used simple motion sensing to remove noise from movement artifacts to improve signal reliability in ambulatory environments. Yan et al.\cite{yan2008efficient} used a more sophisticated feature extraction to remove motion and other noise artifacts in the context of at home fingertip pulse oximeters used for telehealth monitoring.

Liaqat et al. are currently working on using wrist-worn devices to aid COPD patients in treatment and disease management in the context of the WearCOPD project \cite{Liaqat:2016:PWM:2938559.2938606}. Although they currently do not employ \({SpO_2}\) in their consideration of patient health, this project could aid their work by providing a reliability measure for \({SpO_2}\) readings.

\section{Conclusion}
\label{conclusions}

In this work we study the reliability of \spo{} measurements from a wrist-worn pulse oximeter, and show that existing algorithms do not provide reliable readings.
We propose \codename, which uses automated feature extraction and statistical machine learning to identify reliable peripheral oxygen saturation readings taken from the wrist.
After pruning unreliable results with \codename, we show that we can reduce error in the measurements taken from the wrist by up to an order of magnitude. Additionally we demonstrate that even after pruning results, the frequency of reliable readings is still high enough to be useful, and in fact significantly better than current methods that require user intervention. We discuss the effects of sensor placement and skin tone on \codename, explore the effect of IMU information, and propose platforms for user level calibration. 
Finally, we discuss next steps to deploy this technique in the wild. We present this research as a proof of concept for manufacturers and developers to implement reliable data collection platforms with which to build useful applications based on peripheral oxygen saturation.


\bibliographystyle{authordate1}
\bibliography{ms}

\end{document}